\documentclass[12pt,thmsa]{article}
\usepackage{amssymb}
\usepackage{graphicx}
\makeatletter
\newcommand{\row}[1]%
{\mathord{\buildrel{\lower3pt%
\hbox{$\scriptscriptstyle\rightarrow$}}\over #1}}

\newcommand{\dyadic}[1]{\mathord{\dyadic@rrow{#1}}}
\newcommand{\dyadic@rrow}[1]{
\begin{picture}(12,12)(-1,0)
\put(-2,10){\makebox(0,0)[t]{$\scriptscriptstyle\downarrow$}}
\put(-2,11){\makebox(0,0)[l]{$\scriptscriptstyle\longrightarrow$}}
\put(5,0){\makebox(0,0)[b]{$#1$}}
\end{picture}
}

\newcommand{\bra}[1]{\bigl\langle #1 \bigr|}
\newcommand{\ket}[1]{\bigl| #1 \bigr\rangle}

\input{tcilatex}
\begin{document}

\begin{center}
\textbf{Maximum entangled states and quantum Teleportation via
Single Cooper pair Box }

N. Metwally$^{1}$ and A. A. El-Amin$^{2}$  \\
$^{1}$ Math. Dept., College of Science,
Bahrain University, 32038 Bahrain.
$^{2}$Phys. Dept., Faculty  of Science, South valley
University, Aswan, Egypt.\\

\end{center}

\textbf{{Abstract}: } In this contribution, we study  a single
Cooper pair box interacts with a single cavity mode.  We show the
roles   played by the detuning parameters and charged capacities
on the degree of entanglement. For large values of the detuning
parameter the survival entanglement increases on the expanse of
the degree of entanglement. We generate a maximum entangled state
and use it  to perform the original teleportation protocol. The
fidelity of the teleportated state is increased with decreasing
the detuning parameter and  the number of photon inside the
cavity.


\section{Introduction}
Several schemes have been proposed for implementing quantum
computer hardware in solid state quantum electronics. These
schemes use electric  charge \cite{bib1}, magnetic flux\cite
{bib2} and superconducting phase \cite{bib3} and  electron spin
\cite{bib4}. The basic element of the quantum information is the
quantum bit (qubit) which is considered as a two level system.
Consequently, most of the research concentrate to generate
entanglement between two level systems \cite{Zhang}. Among these
systems, the Cooper  charged pairs, due to its properties as a two
-level quantum system, which makes it a candidate  as a qubit in a
quantum computer \cite{Mak, Wend}. So there are a lot of studies
have  been done on these particle from different points of view in
the context of quantum information. One of the most important
tasks in the context of quantum information, is the quantum
teleportation, which emerges from the quantum entanglement, and
 since the first quantum teleportation protocol was introduced by
 Bennett et al \cite{ben}, there are a lot of attentions has been
payed to it \cite{Cabello, Cao, Bou}.

In our contribution, we consider a system consists of a single
Cooper pair interactsa with a cavity mode. The
 separability problem is  investigated, where  the intervals of time
in which the generated state is entangled or separable are
determined. On the other hand under a specific circumstance one
can use this system to generated a maximum entangled state.
Finally we use the generated entangled state to perform the
quantum teleportation.

The paper is organized as follows:  The description of the system
and its solution are introduced in Sec.$2$.  In Sec.$3$,  the
separability problem and the degree of entanglement contained in
the generated entangled state are investigated.  Sec.$4$, is
devoted to  study the effect of the field and the charged qubit
parameters on the phenomena of entanglement and quantum
teleportation.

\section{The Model  and its evolvement}
The single superconducting charged qubit consists of a small
superconducting island with Cooper pair charge $Q$. This island
connected by two identical Josephson junctions, with capacitance
$C_j$ and Josephson coupling energy $E_j$, to  a superconducting
electrode \cite {Mig,Pas}.  This system is described by the
Hamiltonian,
\begin{equation}\label{1class}
H_s=4E_c(n-n_g)^2-E_j\cos\phi,
\end{equation}
where $E_c=e^2/2(C_g+C_j)$ is the charging energy, $E_j=\hbar
I_c/2e$ is the Josephson  coupling energy, $e$ is the charge of
the electron, $n_g=C_gV_g/2e $ is the dimensionless gate charge,
$C_g$ is the gate capacitance, $v_g$ is controllable gate voltage,
$ n$ is number operator of excess cooper pair on the island and
$\phi$ is phase operator \cite{Mig}.

The Hamiltonian of the system (\ref{1class}) can be simplified to
a very simple form, if the Josephon coupling energy $E_j$ is much
smaller than the charging energy i.e $E_j<< E_c$. In this case,
the Hamiltonian of the system can be parameterized by the number
of Cooper pairs $n$ on the island. If  the  temperature is low
enough, the system can be reduced to two-state system (qubit)
controlled by \cite{Sch, You},
\begin{equation}
H_s=-\frac{1}{2}B_z\sigma_z-\frac{1}{2}B_x\sigma_x,
\end{equation}
where $B_z=E_{cl}(1-2n_j) $, $E_{cl}$ is the electric energy and
$B_x=E_j$ and $\sigma_z$, $\sigma_z$ are Pauli matrices.
 This Cooper pair can be viewed as  an atoms with large dipole moment
  coupled to microwave frequency photons in
a quasi-one-dimensional transmission line cavity (a coplanar
waveguide resonator). The combined Hamiltonian for qubit and
transmission line cavity  is given by,

\begin{equation}\label{ham1}
H=\omega a^{\dagger} a+\omega_c\sigma_z-\lambda(\mu-\cos
\theta\sigma_z+\sin\theta\sigma_x)(a^{\dagger}+a),
\end{equation}
where $\omega$ is the cavity resonance frequency,
$\omega_c=\sqrt{E^2_j+[4E_c(1-2n_g)]^2}$ is the transition
frequency of the Cooper pair qubit, $\sigma_z$ and $\sigma_x$ are
Pauli matrices,
$\lambda=\frac{\sqrt{C_j}}{C_g+C_j}\sqrt{\frac{e^2\omega}{2
\hbar}}$, is coupling strength of resonator to the cooper pair
qubit, $\mu=1-n_g$, $\theta=arctan\{\frac{E_j}{E_c(1-2n_g)}\}$, is
the mixing angle.

Assume that we consider the charge degeneracy  point, i.e
$n_g=\frac{1}{2}$ and the radiation  quantized field  is weak. In
this case we can neglect the fast oscillation by using the
rotating wave approximation. Then the Hamiltonian (\ref{ham1})
takes the form
\begin{equation}
H=\omega a^{\dagger}
a+\frac{1}{2}\omega_c\sigma_z-\lambda(a^{\dagger}
\sigma_{-}+\sigma_{+}a),
\end{equation}
where $\sigma_{+}$ and $\sigma_{-}$ are the rasing and lowering
operators such that $[\sigma_{+},\sigma_{-}]=\sigma_z$.

To investigate the dynamics of the total system (cooper pair box
and the filed), let us consider that the charged qubit prior to
the interaction, to be prepared in a superposition of its excited
and ground state, i.e
$\ket{\psi_c(0)}=\alpha\ket{g}+\beta\ket{e}$, and the filed is
prepared in the number state $\ket{\psi_f(0)}=\ket{n}$. The time
development of the state vector $\ket{\psi(t)}$ of the system is
postulated to be determined by Schr\"{o}dinger equation
\begin{equation}\label{sch}
i\hbar\frac{d}{dt}\ket{\psi(t)}=H\ket{\psi(t)}
\end{equation}
The solution of Eq.(\ref{sch}) can be written as
$\ket{\psi(t)}=U(t)\ket{\psi(0)}$, where $U(t)$ is the unitary
operator. In  an explicit for the evolvement of the density
operator $\ket{\psi(t)}\bra{\psi(t)}$ is given by
\begin{eqnarray}\label{EqFinal}
\rho(t)&=&A
\Bigl(|\alpha|^2\ket{g,n}\bra{g,n}+|\beta|^2\ket{e,n}\bra{e,n}\Bigr)
+B\alpha\beta^*\ket{g,n}\bra{e,n}+B^*\alpha^*\beta\ket{e,n}\bra{g,n}
\nonumber\\
&+&iC\sqrt{n+1}(\alpha\beta^*\ket{g,n}\bra{g,n+1}+\beta\alpha^*\ket{g,n+1}\bra{g,n}\Bigr)
\nonumber\\
&-&iC\Bigl(\sqrt{n+1}|\beta|^2\ket{e,n}\bra{g,n-1}+\sqrt{n}\beta\alpha^*\ket{e,n}\bra{e,n-1}\Bigr)
\nonumber\\
&+&i\eta
C^*\Big(\sqrt{n+1}|\beta|^2\ket{g,n+1}\bra{e,n}-\sqrt{n}|\alpha|^2\ket{g,n}\bra{e,n-1}\Big)
\nonumber\\
&+&\eta^2S^2_n\sqrt{n}\sqrt{n+1}\Bigl(\beta\alpha^*\ket{g,n+1}\bra{e,n-1}+
\alpha\beta^*\ket{e,n-1}\bra{g,n+1}\Bigr)
\nonumber\\
&+&S^2_n\Bigl(n|\alpha|^2\ket{e,n-1}\bra{e,n-1}+(n+1)|\beta|^2\ket{g,n+1}\bra{g,n+1}\Bigr)
\nonumber\\
&+&i\sqrt{n}S_n\Bigl(|\alpha|^2C\ket{e,n-1}\bra{g,n}+
\alpha\beta^*C^*\ket{e,n-1}\bra{e,n}\Bigr)
\end{eqnarray}
where
\begin{eqnarray}
A&=&C^2_{n+1}+\frac{\Delta^2}{4}S^2_{n+1},~\quad
B=C^2_{n+1}-\frac{\Delta^2}{4}S^2_{n+1}-i\Delta S_{n+1}C_{n+1}
\nonumber\\
 C&=& i S_n(C_{n+1}+\frac{i\Delta}{2}S_{n+1}),~\quad C_n=\cos{(\gamma\mu_n
 \tau)},\quad
 S_n=\frac{\sin(\gamma\mu_n \tau)}{\mu_n},
 \nonumber\\
 \gamma&=&\frac{\sqrt{C_j}}{C_g+C_j},\quad
 \tau=\sqrt{\frac{e^2\omega}{2\bar h}} t,\quad\mbox{ is the scaled
 time}
\end{eqnarray}
with $\mu_n=\sqrt{\frac{\Delta^2}{4}+\lambda n}$,
$\Delta=E_j-\omega$ is the detuning between the Josephson  energy
and the cavity field frequency .
\begin{figure}
\begin{center}
\includegraphics[width=18pc,height=10pc]{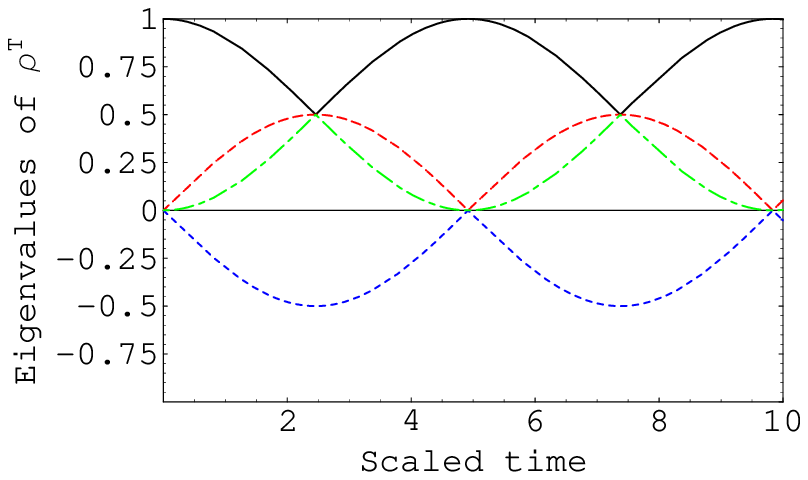}\put(-25,100){$(a)$}
\includegraphics[width=18pc,height=10pc]{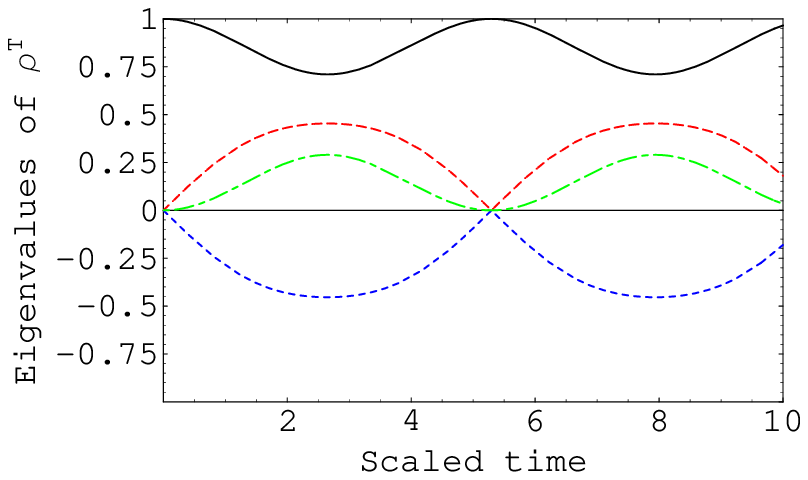}\put(-22,95){$(b)$}\
\includegraphics[width=18pc,height=10pc]{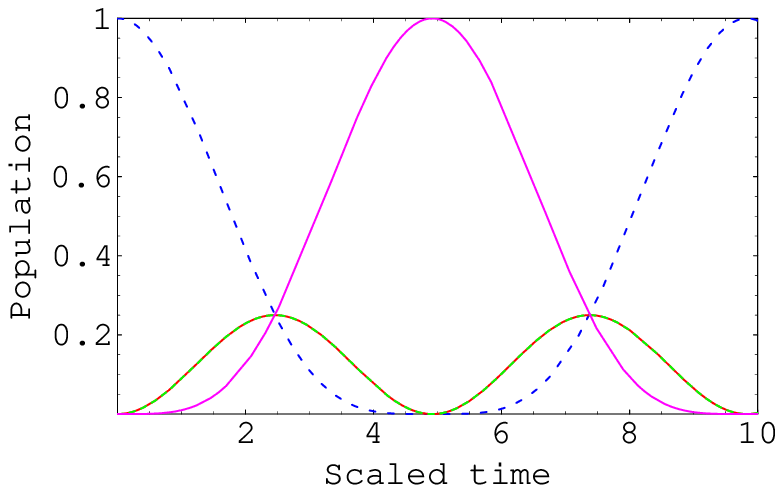}\put(-25,95){$(c)$}
\includegraphics[width=18pc,height=10pc]{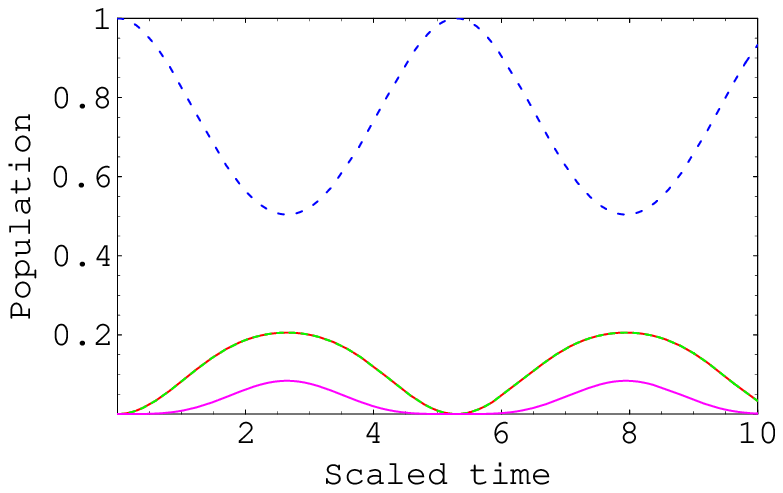}\put(-28,105){$(d)$}\
\includegraphics[width=18pc,height=10pc]{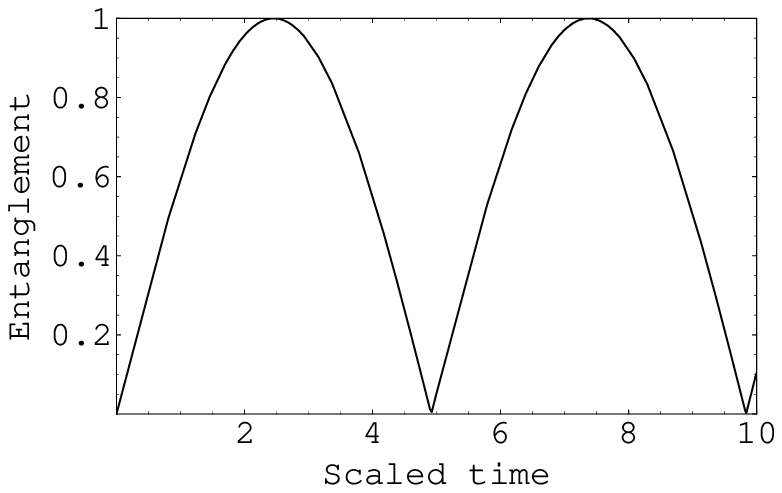}\put(-25,100){$(e)$}
\includegraphics[width=18pc,height=10pc]{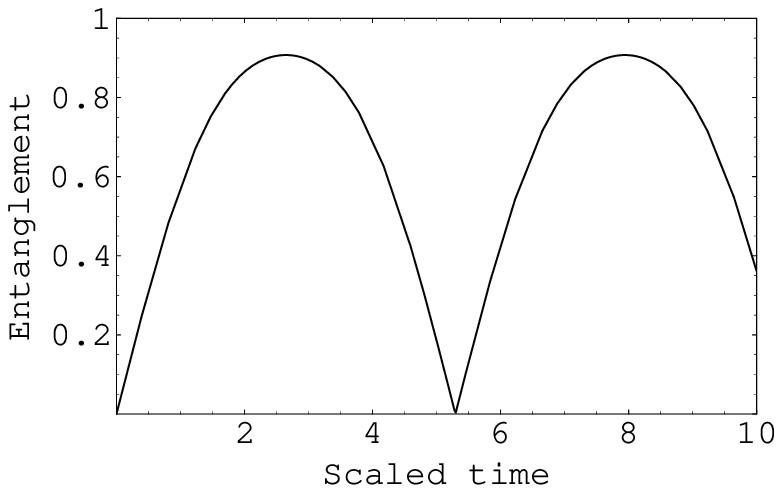}\put(-25,100){$(f)$}
\end{center}
\caption{The behavior of the eigenvalues of the partial transpose
$\rho^T$ is shown in Figs.$(a,b)$. The populations  are shown in
Fig.$(c,d)$ and the degree of entanglement is plotted in
Figs.$(e,f)$. We assume $n=1, C_{jg}=\frac{5}{2}$ and $\Delta =0 $
for Figs.$(a,c,e)$ while for Figs.$(b,d,f)$, $\Delta=1$. }
\end{figure}
\section{Degree of entanglement}
In this section, we study the behavior of the output state
(\ref{EqFinal}) from the separability point of view. To achieve
this task, we plot time evolution of the the eigenvalues of  the
partial transpose eigenvalues of the output density operator
$\rho^{T_2}$ \cite{Peres,Hord}. Also we investigate  time
development  of the occupation probabilities and  the degree of
entanglement which contained in the output state. There are
different  measures  known for quantifying the degree of
entanglement in a bipartite system, such as  the entanglement of
formation \cite{Hill,Benn}, entanglement of distillation
\cite{Woot}, negativity \cite{Eisert}.

 In our calculations we consider the concurrence as a measure of the degree
of entanglement. For two qubits,  the concurrence is calculated in
terms of the eigenvalues $\eta_1,\eta_2,\eta_3$ and $\eta_4$ of
the matrix $R=\rho\sigma_y\otimes
\sigma_y\rho^*\sigma_y\otimes\sigma_y.$ It is given by
\begin{equation}
\mathcal{C}=max\{0, \eta_1-\eta_2-\eta_3-\eta_4\}, \mbox{where}~
 \eta_1\geq\eta_2\geq\eta_3\geq\eta_4.
\end{equation}
  For maximally
entangled states concurrence is $1$ while $\mathcal{C}=0$ for
separable states\cite{Woot}.

In the first example, we assume that the Cooper-pair box is
initially prepared in the ground state and the field is prepared
in the Fock state, i.e the initial state of the system is given by
$\ket{g,n}$.
 In Fig.(1), we investigate  the effect of the detuning parameter on
 the behavior of the eigenvalues of the partial transpose
 of the density operator of the output state, the time evaluation of
 the occupation probabilities   and the degree of entanglement.
 For these numerical calculations,
 we assume that the ratio  between the Josephson junction capacity,
 $C_j$, and the gate capacity $C_g$, is defined by  $C_{jg}=\frac{C_j}{C_g}=\frac{5}{2}$
 and $n=1$.  In Fig.$(1a)$, we  see that as the interaction goes on, i.e the
 scaled time, $\tau>0$, there is only one   negative eigenvalues
and the rest are non-negative. According to the
 Peres-Horodecki criterion \cite{Peres,Hord}, there is an entangled state is
 generated. As the time increases we see that at specific time
 $\tau\simeq 5$ all the eigenvalues are non-negative. This
 means that the entangled qubit turns into a product state
 (separable).
\begin{figure}
\begin{center}
\includegraphics[width=18pc,height=11pc]{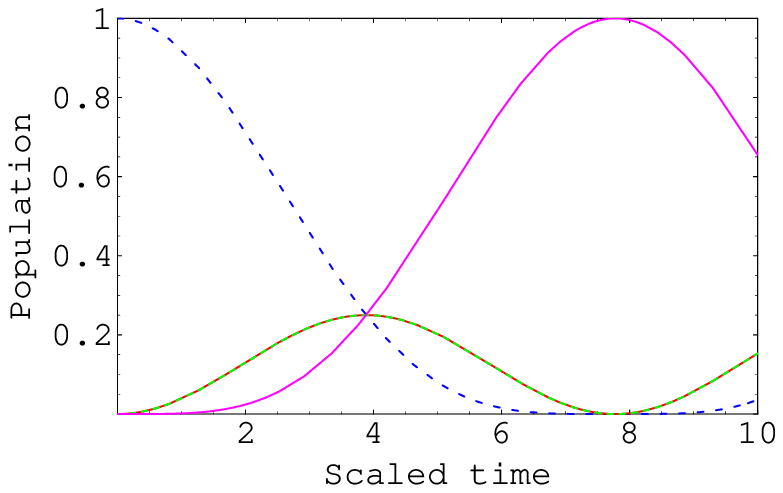}\put(-23,115){(a)}
\includegraphics[width=18pc,height=11pc]{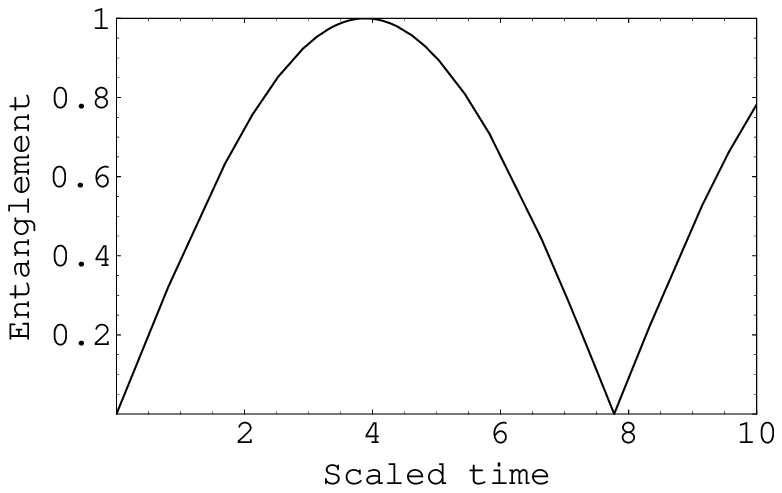}\put(-25,115){(b)}
\end{center}
\caption{ The same as Fig.$(1)$, but $\Delta=0,
C_{jg}=\frac{2}{5}$. }
\end{figure}
In Fig.$(1c)$, we plot the occupation probabilities as functions
of the scaled time. Note that the populations of the four states
exist and it has different values for the diagonal  occupations
but for the off-diagonal occupations are completely coincide. For
this reason  an entangled state  is generated as soon as the
interaction starts. In this case the entangled state is
\begin{equation}\label{ent1}
\rho=\kappa(\ket{n,g}\bra{n-1,e}+\ket{n-1,e}\bra{n,g})+
\chi_1\ket{n,g}\bra{n,g}+\chi_2\ket{n-1,e}\bra{n-1,e},
\end{equation}
where $\chi_1$ and $\chi_2$ are the probability to find the system
in the state $\ket{n,g}\bra{n,g}$  and $\ket{n-1,e}\bra{n-1,e}$,
respectively. In computational basis one can write it as
$\rho=\kappa(\ket{1,0}\bra{0,1}+\ket{0,1}\bra{1,0})+
\chi_1\ket{1,0}\bra{1,0}+\chi_2\ket{0,1}\bra{0,1}$. It is clear
that the first bract is one type of  Bell state
$\ket{\psi^{+}}\bra{\psi^+}$.
 However at $\tau\simeq 2.5$, the partially
entangled state (\ref{ent1}) becomes a maximum, where all the
occupation probabilities  have equal occupation probabilities at
this point. This result is shown in  from Fig.(1e), where we
quantify the degree of entanglement by using the concurrence. From
this figure, we can see  that the degree of entanglement is
increased  as the time increased and reaches to the unity at
$\tau\simeq 2.5$. Given enough time, the system will therefore
reaches a state where all the  occupation probabilities vanish. At
specific time the system turns into a product state, this happens
at $\tau\simeq 5$, (see Fig.$(1a)$). Also, on the left hand side
of Fig.(1), we consider $\Delta=1$ and the other parameters are
fixed, it is clear  that as one increases $\Delta$, the system
turns into a product round $\tau=5.4$. This means that the time of
the entanglement survival  increased. This phenomenon is clear
shown by comparing Fig.(1a) and Fig.(1b). The behavior of the
occupation probability is seen in Fig.(d), where there is no
intersection point between the off diagonal occupation
probabilities and the diagonal ones. So in this case one can not
get a maximum entangled state. As time goes on the  system becomes
a separable at $\tau\simeq 5.4$, this result agrees  with that
depicted in Fig.$(1b)$.  The amount of entanglement contained in
the output state is  shown in Fig.$(1f)$. From this figure, we can
notice that the maximum entanglement is less than unity. So one
can say that by increasing the detuning parameter, one can
increase the survival time of entanglement on the expanse of the
degree of entanglement.

The effect of the ratio of $C_{jg}$ is seen in Fig.(2), where we
consider a small value of this ratio. We investigate the behavior
of  the  occupation probabilities and the degree of entanglement
only.  It is  clear that as one decreases $C_{jg}$, the first
maximum entangled state is obtained round $\tau=4$. This means
that for small ratio, one takes a larger time to generate maximum
entangled state. Also, the point at which the system turns into a
separable state is shifted. The degree of entanglement contained
in this state is shown in Fig.(2b), where  it has a  unity value
at the maximum entangled state. So, the ratio $C_{jg}$ has no
effect on the value of the degree of entanglement.
\begin{figure}
\begin{center}
\includegraphics[width=18pc,height=11pc]{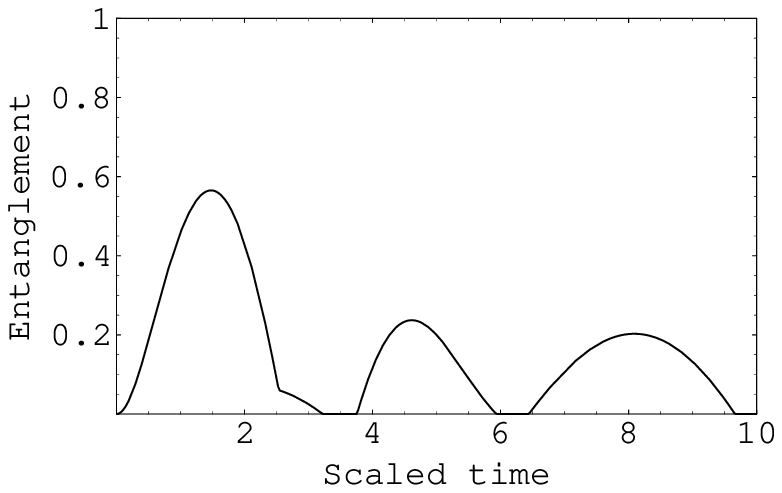}\put(-23,115){(a)}
\includegraphics[width=18pc,height=11pc]{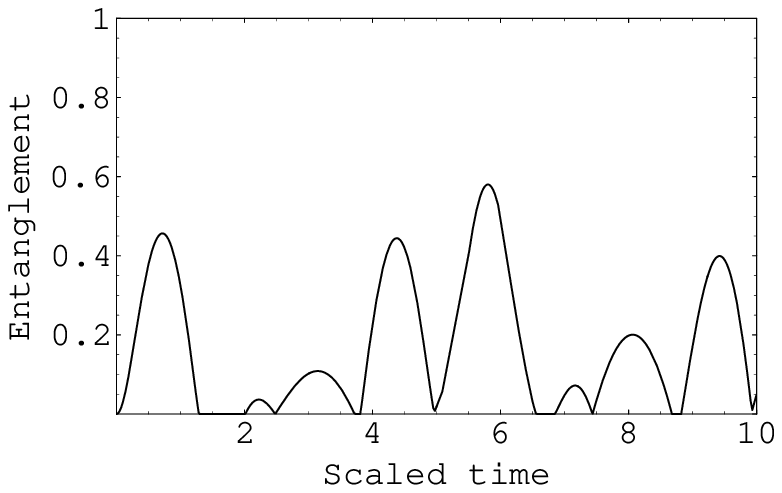}\put(-25,115){(b)}
\end{center}
\caption{ The degree of entanglement for a system is initially
prepared in the superposition state
$\ket{\psi_0}=a\ket{n,g}+(1-a)\ket{n,e}$ with $a=0.5$,
$C_{jg}=\frac{5}{2}$ and (a) $\Delta =0$, (b) $\Delta =1$. }
\end{figure}
\begin{figure}
\begin{center}
\includegraphics[width=18pc,height=12pc]{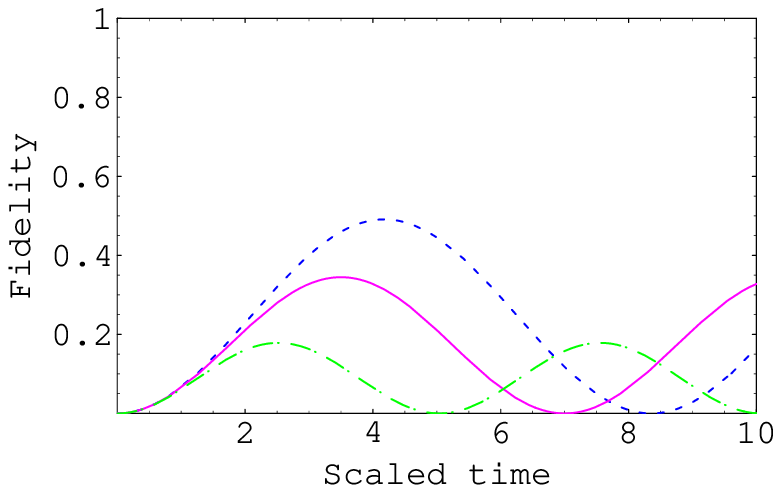}\put(-23,115){(a)}
\includegraphics[width=18pc,height=12pc]{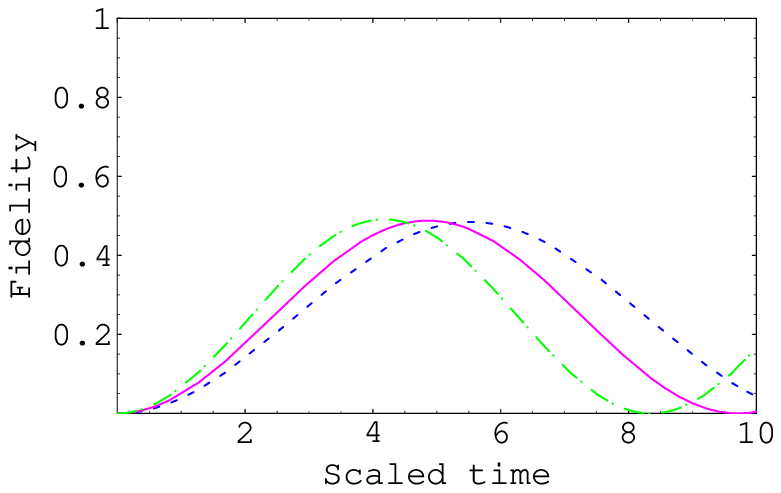}\put(-25,115){(b)}
\end{center}
\caption{ The Fidelity of the teleported state at Bob's hand
(a)For $C_{jg}=5$, number of photon, $n=1$ and dot, solid and the
dashed curves for $\Delta=0.1,0.5$ and 1 respectively (b) The same
as Fig.(a) but $\Delta=0.1$ and the dot, the  solid and the  dash
curves for $C_{jg}=5/3,5/2$ and $5$ respectively. }
\end{figure}
Our second case, we assume that the Cooper pair box is prepared in
the superposition state. So the initially state of the system,
$\ket{\psi_0}=a\ket{n,g}+(1-a)\ket{n,e}$. In Figs.(3), we plot
 the degree of entanglement where we assume that $a=0.5$,
fixed vales of the ratio $C_{jg}=\frac{5}{2}$ and different values
of the detuning parameter $\Delta$. In this case the effect is
completely different ( see Fig.$(3a)$), where the degree of
entanglement is small compared with  by that depicted in
Fig.$(1e)$. Also, the  survival  time of entangled is small
comparing by the previous case, where the system behaves as   a
separable system several times in a small range of time. In
Fig.$(3b)$, we increase the value of the detuning parameter
($\Delta=1$). It is clear  that the instability of the system
increases and both of the sudden death \cite{Eberly} and sudden
birth of entanglement is seen \cite{Ficek}. Now, one can see that
by controlling  the Cooper pair  box parameters and the field
parameters, one can generate entangled state with high degree of
entanglement. One of the best strategy is preparing the Cooper
pair box in the excited or the ground state. If the charged qubit
and the field in a resonance i.e $\Delta=0$, one can generate a
maximum entangled state. By decreasing the ratio $C_{jg}$, the
survival time of entanglement is much larger.

\section{Teleportation}
No, we want to use the generated entangled state to  achieve the
quantum teleportation by  using the output state (\ref{EqFinal}),
as a quantum channel. Assume that Alice is given  unknown state
 defined by
 \begin{equation}\label{ste}
 \ket{\Psi}=\lambda_1\ket{0}+\lambda_2\ket{1},
 \end{equation}
where $\lambda^2_1+\lambda^2_2=1$. She wants to sent this state to
Bob through their quantum channel. To attain this aim, Alice and
Bob shall  use the original teleportation protocol \cite{ben}. In
this case, the total state of the system is
$\rho_{\psi}\otimes\rho^{out}$, where $\psi$ is given by
(\ref{ste}) and $\rho^{out}$ is defined by (\ref{EqFinal}). Alice
makes  measurement on the given qubit and her own qubit. Then she
sends her results through a classical channel to Bob. As soon as
Bob receives the classical data, he performs a suitable unitary
operation on his qubit to get  the teleported state.
\begin{figure}
\begin{center}
\includegraphics[width=25pc,height=12pc]{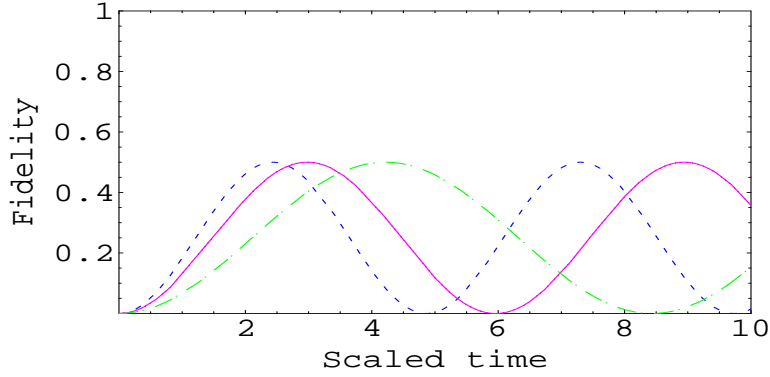}
\end{center}
\caption{The same as Fig.$4$, with $C_j/C_g=5$ and $\Delta=0.1$
where the dot, the solid and the dash curves for $n=3,2,1$
respectively.}
\end{figure}
Let us assume that Alice measures $\ket{\psi^+}$, then the density
operator on Bob's hand is given by
\begin{equation}\label{Bob}
\rho_{Bob}=\frac{1}{2}(|\lambda_1|^2|B_n|^2\ket{0}\bra{0}+\lambda_1\lambda_2^*B_n
A_n^*\ket{0}\bra{1} +\lambda_2^*\lambda_1
A_nB_n^*+|\lambda_2|^2|A_n|^2\ket{1}\bra{1}),
\end{equation}
with,
\begin{equation}
A_n=\cos(\gamma\mu_n\tau)-\frac{i\Delta}{\mu_n}\sin(\gamma\mu_n\tau),
\quad B_n=\frac{i\lambda\sqrt{n}}{\mu_n}\sin(\mu_n t)
\end{equation}
where, we  assume that the  Cooper pair box  is prepared on the
ground state and the field on the Fock state i.e
$\ket{\psi_0}=\ket{g,n}$. In Fig.4, we plot the of the fidelity of
the teleported  state (\ref{Bob}), for different values of the
field and the Cooper pair parameters. The effect of the detuning
parameter is seen in Fig.(4a), where we assume that there is only
one photon inside the cavity and the ratio $C_{jg}=5$. From this
figure it is clear that as one increases $\Delta$, the fidelity of
the teleported state decreases. This due to that for large values
of the detuning parameter, the degree of entanglement decreases as
it is clear from Fig.$(1f)$. Since, the entangled time, the time
in which the entanglement survival,  increases  for small values
of the detunang parameter, the fidelity vanishes for large values
of $\Delta$ much faster. The effect of  $C_{jg}$ is shown in
Fig.(4b),where for small values of this ratio, the fidelity
reaches to its maximum value faster than the large values of
$C_{jg}$. On the other hand the different values of this ratio has
no effect on the maximum values of the fidelity. Also, as one
increases this ratio, the interval of time in which the channel is
available for quantum teleportation increases.

In Fig.(5), we investigate the effect of different values of the
number of photons inside the cavity, where we fixed the other
parameters. It is clear that for small values of $n$, the
possibility of sending information with non-vanishing fidelity  by
using the quantum teleportation increases. This is due to that for
large values of $n$, the possibility of quick interaction
increases and consequently one can gets an entangled state much
faster.

\section{Conclusion}
We employ the Cooper pair  box to generate entangled state by
interacting with a single cavity mode is initially prepared in the
numbers state. We show that it is possible to generate a maximum
entangled state by controlling  on the capacities  and the
detuning parameters. Also,  one can use the generated entangled
state to perform the quantum teleportation. We investigate  the
effect of the detuning parameter, the ratio of capacities and  the
photon inside the cavity on the fidelity of the teleported state.
For small values of the detuning parameter and the photon numbers
inside the cavity one can teleportate the given state with large
fidelity. On the other hand by increasing the ratio of the
capacities one can enlarge the interval of time in which one can
teleportate the state with non-vanishing fidelity.


\begin{thebibliography}{99}
\bibitem{bib1}
D. V. Averin, Solid State Communications, {\bf 105} 659 (1998); Y.
Makhlin, G.Schon, and A. Shnirman, Nature {\bf 398} 305 (1999).

\bibitem{bib2}
 J. Mooij, T. Orlando, L. Levitov, L. Tian, C. H. van der
Wal, and S. Lloyd, Science {\bf 285} 1036 (1999).

\bibitem{bib3}
 A. Blais,  A.  Zagoskin
Phy. Rev. A, {\bf 61}, 042308 (2000).
\bibitem{bib4}
D. Loss, D. DiVincenzo, Phys. Rev. A {\bf 57}, 120 (1998).

\bibitem{Zhang}  M. Zhang, J. Zpu and B. Shao, Int. J. Mod. Phys.
B{\bf 16} 4767 (2002), W. Krech and T. Wanger, Phys. Lett A {\bf
275} 159 (2000), H. S. Ding, S. P. Zhao, G. H. Chen and Q. S.
Yang, Physica C{\bf 382} ~431 (2002).
\bibitem{Mak}
Yu. Makhlin, G. Sch\"{o}n and A. Shnirman, Nature {\bf 398}~305
(1999); G. Sch\"{o}n, Yu. Makhlin and A. Shnirman, Physica C{\bf
352} ~113(2001); J. Q. You and F. Nori, Physica E {\bf 18} 33
(2003).
\bibitem{Wend}
G.  Wendin, R. Soc. Lond. A {\bf 361} 1323 (2003), A. Niskanen, J.
Vartiainen, and M. Salomaa, Phys. Rev. Lett.{\bf 90}  197901
(2003).
\bibitem{ben}
C. H. Bennett, G. Brassard, C. Crepeau, R. Jozsa, A. Peres, and W.
Wootters, Phys. Rev. Lett. 70, 1895 (1993).

\bibitem{Cabello}
A. Cabello, Phys. Rev. Lett {\bf 89}, 100402 (2003); B. Zeng, X.
S. Liu, S. Y. Li and G. L. Long, Commun. Theor Phys. {\bf 38} 537
(2003).

\bibitem{Cao}
H,-J. Cao, H. C. Zhong and S. He-Shan, Phys. Scr {\bf 78} 015002
(2008).
\bibitem{Bou}
D. Bouwmeester, J.-W. Pan, K.Mattle,M. Eible, H.Weinfurter, and A.
Zeilinger, Nature {\bf 390}, 575 (1997); D. Boschi, S. Branca, F.
D. Martini, L. Hardy, and S. Popescu, Phys. Rev. Lett.{\bf 80}
1121 (1998).

\bibitem{Mig}
R. Migliore, A. Messina, A. Napoli, Eur. Phys. J. B {\bf13} 585
(2000);R. Migliore, A. Messina, A. Napoli, Eur. Phys. J. B {\bf
22} 111 (2001).
\bibitem{Sch}
Y. Makhlin, G. Schon and A. Shnirman, Rev. Mod. Phys. {\bf 73} 357
(2001).
\bibitem{Pas}
T. S. Tsai, Y. Nakamura and Yu. Pashkin, Physica C {\bf 367}191
(2002);
 Yu. A. Pashkin, T, Yamamoto, O. Astafiev, Y. Nakamura, D,
Averin, T. Tilma, F. Nori and J. S. Tsai, Physica C {\bf 426}1552
(2005).
\bibitem{You}
J. You, F.Nori, Physica E{\bf 18} 33 (2003).
\bibitem{Peres}
A. Peres, Phys. Rev. Lett. A {\bf 77}, 1413 (1996).
\bibitem{Hord} M. Horodecki, P. Horodecki, R. Horodecki, Phys.
Lett.A {222} 1 (1996).

\bibitem{Hill}
S. Hill, W. K. Wootters, Phys. Rev. Lett.,{\bf 78} 5022 (1997).
\bibitem{Benn}
C. H. Bennett, H. J. Bernstein, S. Popescu, B. Schummacher, Phys.
Rev. A {\bf 53}2046 (1996).

\bibitem{Woot}
W. K.  Wootters, Phys. Rev. Lett.{\bf 80} 2245 (1998).

\bibitem{Eisert}
P. Horodecki, Phys. Lett A{\bf 232} 333(1997); J. Eisert, M. B.
Plenio, J, Mod. Opt. {\bf 46}145 (1999); J. Lee, M. S. Kim, Y. J.
Park, S. Lee, J. Mod. Opt. {\bf 47} 2151 (2000).
\bibitem{Eberly}
T. Yu and J. H. Eberly, Opt. Commu.{\bf 264}~{393}~{(2006)}; M.
Yonac, T. Yu and J. H. Eberly, J. Phys. B{\bf 39}~{S621}~{(2006)};
T Yu and J. H. Eberly; J. Mod. Opt. 54, 2289-2296 (2007).

\bibitem{Ficek}  Z. Ficek and  R. Tanas, quant/ph  :08024287
(2008); M. Abdel-Aty and T. Yu, quant/ph 0805.3576 (2008).



\end{thebibliography}
\end{document}